# Nitrogenated CQD decorated ZnO nanorods towards rapid photodegradation of rhodamine B: a combined experimental and theoretical approach

Sujoy Kumar Mandal[1], Sumana Paul[2], Sujoy Datta[1], Debnarayan Jana[1*]

[1]Department of Physics, University of Calcutta, 92, A.P.C. Road, Kolkata-700009, India.

[2]Department of Physics, Indian Institute of Technology Guwahati, Guwahati- 781039, India.

*E-mail: djphy@caluniv.ac.in

**Abstract:**

In this work, hybridization of nonmetal nitrogen-doped carbon dots (NCQDs) with ZnO nanorods (NRs) is utilized towards better photocatalytic degradation of rhodamine B under ultraviolet (UV) irradiation. The structural characterization is confirmed by XRD, XPS, FTIR and HRTEM measurements, while, optical properties have been investigated using UV-visible absorbance spectroscopy and photoluminescence study. The dye degradation using ZnO NR is recorded as ~69%, while the performance of ZnO/NCQD climbs up to ~90%, after nine minutes of UV irradiation. Furthermore, the reusability test suggests better stability of ZnO/NCQD than bare ZnO under photocorrosion. The cyclic voltammetry study confirms that the photoinduced electron-hole pairs originate from the heterojunction established between the interfaces of NCQD and ZnO. An insight on the photocatalytic excellence of ZnO/NCQD system is drawn from the density functional theory study. This indicates that appropriate band alignment of the heterostructure constituents is the key factor in this experimental attempt towards environment remedy.

# 1. Introduction

Following the growth of human awareness of environmental remedies, photocatalysis science has come into focus [1, 2]. While the initial boom has been based on single-component materials, heterostructures are eventually proved to be better alternatives. However, fast recombination of photoexcited charge carriers, and, low surface catalytic reaction efficiency produces a hindrance to the performance of many photocatalytic materials [3]. A series of initial efforts in this field has proved that a single component semiconductor photocatalyst cannot satisfy all the required criteria. Several strategies have been developed to overcome these drawbacks such as bandgap engineering and construction of heterostructure [4, 5]. A detailed study in this avenue has revealed that the proper energy band alignments across the interface of two semiconductors can effectively improve the charge separation by crippling the recombination of the photogenerated electron-hole pairs [6]. Depending upon the relative band alignments of two different semiconductors, nanoscale heterostructure can be classified as type-I, type-II, or type-III. However, semiconductor type II heterostructures, and, the heterostructures with multiple components can be particularly advantageous because of the enhanced photogenerated electron-hole separation rates [7]. In the case of the type-II junction, photoexcited electrons can travel from the semiconductor 1 (SM1) with a more negative conduction band (CB) to the semiconductor 2 (SM2) with less negative CB. Simultaneously, photogenerated holes can travel in the opposite direction; from the more positive valence band (VB) of SM1 to less positive VB of SM2 [8, 9]. This leads to a complete separation of photogenerated electron-holes in two different semiconductors, thus, enhancing the photocatalytic performance.

In the last decade, various inorganic semiconductors like $TiO_2$, ZnO, ZnS, and $MoS_2$ have mostly been used for the removal of toxic pollutants from the environment due to their efficiency and broad applicability [7, 10-16]. Amongst, zinc oxide (ZnO) has been utilized for solar energy conversion and environmental applications because of its low toxicity, high electron exchange performance, earth abundance, high photostability, capability of rapid electron-hole pair generation, and, high efficiency towards photocatalysis [7, 17].

Also, various nanostructures of ZnO with desirable surface to volume ratio can be easily synthesized, and, electrical and optical properties of those are easily tunable by various types of doping or heterojunction formation. Among them, 1D heterostructures have gained considerable interest due to their easy charge transport properties. Such migration of charge carriers requires a suitable potential gradient from one component to the other, which is strongly related to the morphology, surface modification and the nature of the heterostructures [18]. Doping with various metals and non-metals [19-21], and, surface modification with novel metals [22, 23] play important role as well. In recent times, heterostructure formation with different nanomaterials [24, 25], e.g., graphene oxide, reduced graphene oxide and carbon nanotubes [26-29] with ZnO are extensively studied as potential photocatalyst for atmospheric contaminant removal, antibacterial properties as well as for photocatalytic energy production. However, there is still enough space for exploration of the prospect of carbon quantum dot (CQD)-based ZnO heterostructures as a potential photocatalyst.

In recent times, CQD based heterogeneous photocatalysts are drawing some attention. CQD is a newly found quasi-zero-dimensional nanomaterial exhibiting distinct properties including large two-photon absorption cross-section, wide absorption band, high optical absorption coefficient, chemical stability and excellent biocompatibility [30]. Besides having unique upconversion photoluminescence property, it also plays the role of an electron acceptor to reduce electron-hole pair recombination [31]. All these features have established CQD as an excellent material to design heterostructures in search of an efficient photocatalyst. On the other hand, quantum dots can serve the purpose of another key factor to reinforce photocatalytic activity, the trapping and transferring of the electrons, [32]. Besides, CQD prevents the photocorrosion of ZnO by its shielding nature [33].

Recently, Pirsaheb et. al. have discussed in their review article about various CQD-based heterostructures and their use in photocatalysis [34]. For example, CQD/ZnO nanoflower composites synthesized by a two-step method have exhibited photocatalytic excellence towards the degradation of rhodamine B (RhB) dye [35]. Muthulingam et. al. have synthesized CQD/N-ZnO composite and Song et al. have synthesized N-P co-doped CQD decorated multishell ZnO as a photocatalyst [36, 37]. Now, N doping in CQD has a lot of advantages in enhancing its optoelectronic properties [38, 39]. The atomic size of N is comparable to C, making it a common choice for doping CQDs. Even it has been elucidated that N-doped CQD can efficiently induce charge delocalization, lower down the work function of nanodots, and, also effectively enhance the electron transferability [40, 41]. So, nitrogen doping can induce new unique physical and chemical properties into carbon dots. Consequently, heterostructures with CQDs lead to superior photocatalytic activity [42]. Till now, very few investigations have been carried out to couple N-doped CQD with semiconductors like $TiO_2$ [41, 43] and g-$C_3N_4$ [44] to increase their photocatalytic activity. Moreover, photocatalysis being a surface phenomenon, so one-dimensional ZnO nanostructure is essentially effective for improved photocatalytic behavior due to its high surface-to-volume ratio. Motivated by the pivotal features of CQD heterostructures with several other semiconductors and advantages of N-doped CQD over CQD, herein, we have synthesized N-doped CQD (NCQD) decorated ZnO nanorods via a soft chemical route. The structure, morphology, optical and photoelectric properties have been explored in detail. The photocatalytic activity of ZnO/NCQD heterostructures has been evaluated by the degradation of RhB dye under UV irradiation. The relationship between the structure and photocatalytic activity has been investigated through multiple techniques. A theoretical density functional theory (DFT) based insight regarding the mechanism of action of NCQDs for the enhanced photocatalytic behaviour has been complemented.

## 2. Experimental Section

### 2.1 Materials

All chemicals are of analytical grade and have been used without further purification. Zinc nitrate hexahydrate ($ZnNO_3 \cdot 6H_2O$), ethanolamine and hydrogen peroxide ($H_2O_2$) have been purchased from Merck. Hexamethylenetetramine (HMT), polyvinylpropeline (PVP) and rhodamine B have been purchased from Sigma-Aldrich. Deionized water (DI) has been used for synthesis purposes.

### 2.2 Synthesis of ZnO nanorods

ZnO nanorods (NRs) have been synthesized via a sol-gel method [45]. Typically, 0.5 g PVP has been dispersed in 50 ml DI water containing in a 500 ml beaker, and then, has been inserted in an oil bath at 90 °C under vigorous stirring. Two separate solutions have been made by dissolving 1.78 g $ZnNO_3 \cdot 6H_2O$ in 50 ml DI water and 0.280 g HMT in another 50 ml DI water. Then, $ZnNO_3 \cdot 6H_2O$ solution has been dropwise added to the PVP-water mixture and after that HMT solution has been dropwise annexed to this solution. After a few minutes, the solution has turned whitish and kept at the same temperature (90 °C) under the vigorous stirring condition for 3 hr. After cooling to ambient temperature, the solution has been centrifuged several times using an ethanol-water mixture to remove excess PVP and unwanted chemicals from the ZnO surface. Finally, the centrifuge tube containing the white precipitate has been dried in an oven overnight at 70 °C. In the end, the synthesized ZnO has been collected by mortaring.

### 2.3 Synthesis of nitrogen-doped carbon quantum dot

Nitrogen-doped carbon quantum dots have been synthesized by the pyrolysis of ethanolamine [46]. In brief, 3 ml of ethanolamine has been mixed with 4.5 ml of $H_2O_2$ in a three-neck vial of 25 ml capacity and heated at 170 °C for 30 min. After that, this solution has been collected and diluted by adding 100 ml of DI water. This solution has been treated as a stock one for NCQD and further used for the synthesis of the composites.

### 2.4 Synthesis of nitrogen-doped CQD decorated ZnO nanorods

Again, NCQD decorated ZnO nanorods have been synthesized in an oil bath at 90 °C. Typically, 40 mg ZnO NR has been suspended in 40 ml DI water contained in a 250 ml beaker, followed by a 10 min sonication. After that, required amount of NCQD solution has been dropwise added under vigorous stirring and the solution has been kept at 90 °C for 3 hr with stirring at 500 rpm. Finally, NCQD decorated ZnO NRs have been collected by centrifugation as described in the case of ZnO NR synthesis. The collected composites have been dried in an oven overnight at 70 °C. Based on the addition of 200, 400, 600 μl NCQD aqueous solution from the stock, the synthesized composites have been named as ZnO/NCQD-1, ZnO/NCQD-2 and ZnO/NCQD-3, respectively.

### 2.5 Characterization

The powder X-ray diffraction (XRD) patterns have been recorded on a Bruker D8 diffractometer equipped with Cu-K$_\alpha$ radiation ($\lambda$=1.5406 Å) operating at 40 kV and 40 mA to study the crystal structure of the synthesized samples. The continuous scans have been collected over a 2θ range of 20° to 80° with a scan rate of 0.03° sec$^{-1}$. The morphologies of the samples have been studied by a high-resolution transmission electron microscope (HRTEM) [UHR-FEG TEM, JEM-2100F, Jeol, Japan] operating at 200 kV. The elemental composition and X-ray mapping have been investigated through energy dispersive X-ray (EDX) measurements (Sigma, Zeiss). X-ray Photoelectron Spectroscopy (XPS) measurements have been carried out with an Omicron Multiprobe (Omicron Nanotechnology GmbH) photoelectron spectrometer fitted with an EA125 hemispherical analyzer, and, a monochromatized Al K$_\alpha$ (hv: 1486.6 eV) source. The identification of functional groups has been performed via FTIR spectroscopy using an FTIR spectrometer (Spectrum One, Perkin Elmer, spectral range 400–4000 cm$^{-1}$, spectral resolution 4 cm$^{-1}$, transmittance mode, KBr pellets). UV-vis absorption spectra of the samples have been recorded using Shimadzu UV 2450 spectrophotometer. Photoluminescence (PL) spectra of all the samples have been documented using Horiba FL 1000 fluorescence spectrometer using 300 nm excitation.

### 2.6 Photocatalytic experiments

The photocatalytic activity of as-synthesized materials has been evaluated by discoloration of RhB dye as a model dye at room temperature. In a typical reaction, 5 mg of the catalyst has been dispersed in a 25 mL aqueous solution of RhB dye (1 x 10$^{-5}$ M), and, sonicated for 10 min. A 70 W UV lamp has been used as an irradiation source and positioned 10 cm above the surface of the reaction solution. In every experiment, before irradiation, catalysts have been suspended into RhB aqueous solution and stirred in dark for 30 min to achieve the adsorption-desorption equilibrium. 2 ml of the suspension has been taken out every 3 min and then centrifuged at 8000 rpm for 5 min to remove the photocatalyst. The concentrations of RhB have been analyzed by a UV–vis spectrophotometer (make: Shimadzu model: UV 2450) according to its absorbance maxima at 553 nm.

### 2.7 Cyclic voltammetry study

The cyclic voltammograms (CVs) of as-synthesized samples have been recorded at 100 mV/sec within the potential range of -2.0 to 2.0 V vs. Ag/AgCl. The working electrode was a glassy carbon disk (0.32 cm$^2$), which have been polished with alumina solution, washed with absolute acetone, and air-dried before each electrochemical run. The reference electrode was Ag/AgCl, with platinum as a counter electrode. The samples have been prepared by dissolving 1 mg photocatalyst in 2ml of DI water followed by 10 min sonication. The CV data have been taken in the suspension method, and, potassium nitrate of 1 M solution has been used as an electrolyte to control the diffusion of the ions. It is to be noted that all the experiments have been performed in standard electrochemical cells at 25 °C.

### 2.8 Photocurrent measurement

The photocurrent measurements of ZnO NR and ZnO/NCQD-2 have been investigated in this study. Before photocurrent measurement, thin films of ZnO NR and ZnO/NCQD-2 have been made on ITO coated PET substrate. 2 mg of photocatalyst has been dispersed in 5 ml of ethanol solution and sonicated for 20 min. Thin films have been made using a spin coater. One drop of the solution has been kept on a PET substrate, and, have been rotated at 4000 rpm for 2 min. After drying under tungsten bulb (100W), again one drop has been added, and, has been rotated for another 2 min. This process has been continued 20 times until the desired thickness is achieved. Finally, silver have been evaporated on the material surface covered with a metallic sheet containing some small circular openings. So, silver and ITO act as electrodes for photocurrent measurement. A UV torchlight (λ=360 nm) has been used as an irradiation source and kept at 10 cm above the catalyst surface.

## 3. Computational methodology

The Density Functional Theoretical (DFT) calculations have been carried out using Quantum Espresso [47] suite. The Projected Augmented Wave (PAW) basis set and Perdew Burke Ernzerhof (PBE) [48] exchange-correlation functional have been used for structural relaxation. For band-gap, work-function and band alignment calculations, PBE0 [49] hybrid functional and norm-conserving pseudopotentials have been used. We have set the kinetic energy cutoff for wavefunctions at 50 Ry for all calculations, and, the kinetic energy cutoff for charge density has been set at 400 Ry for PAW basis calculations. For ZnO 8x8x8 k-points mesh has been used, and for quantum dots, Γ point calculations have been done. We have created a vacuum of 10 Å for quantum dots. To plot the PBE0 bands Wannierization using Wannier90 [50] package has been employed.

## 4. Results and Discussion

### 4.1 XRD analysis

The structural characterization of as-prepared ZnO NR and ZnO/NCQD hybrid composites have been performed by XRD analysis and is presented in Figure 1(a). The peak positions in the patterns for ZnO NR, ZnO/NCQD-1, ZnO/NCQD-2 and ZnO/NCQD-3 have been identified with the standard wurtzite phase of ZnO (JCPDS card no. 79.0207). All ZnO/NCQD composites exhibit the same characteristic diffraction nature of pure ZnO. The major peaks of pure ZnO NR at 2θ = 31.88°, 34.55° and 36.37° correspond to the (101), (002) and (101) planes of wurtzite ZnO. In the case of all the composites, all the peaks are slightly shifted to a lower value than pure ZnO NR. The enlarged image of the most intense peak ((101) plane) is shown in the inset of Figure 1(b). The nominal shifting of the peaks are resulting from the slight change in the crystalline structures of the composites. This is due to the surface hybridization of ZnO with NCQD. Moreover, no obvious characteristic peaks of NCQDs can be detected, as NCQDs in ZnO/NCQD composites are found in small amount, high dispersive and low crystalline form. Such observation is common in other similar systems [40, 46]. The average crystalline size has been estimated using Scherrer's formula, $D = \frac{0.9\lambda}{\beta Cos\theta}$, where $D$ is the average crystalline size, $\lambda$ is the X-Ray wavelength ($k_\alpha$), $\beta$ is the full width at half maxima (FWHM) of the most intense peak corresponding (101) plane, and, $\theta$ is the diffraction angle. The calculated crystalline size of ZnO NR, ZnO/NCQD-1, ZnO/NCQD-2 and ZnO/NCQD-3 are ~ (41.7 ± 0.7), (37.2 ± 0.6), (38.8 ± 0.8) and (39.7 ± 0.8) nm, respectively. Results indicate that no obvious change in crystalline size has occurred after the formation of heterostructure with NCQD. Figure 1(b) displays the XRD pattern of the synthesized NCQD. The broadening of the peak corresponding (002) plane at 2θ ~ 22.7º can be attributed to the intrinsic amorphous nature of NCQD [51]. This is probably the reason for not showing NCQD characteristics distinctly in the composite form.

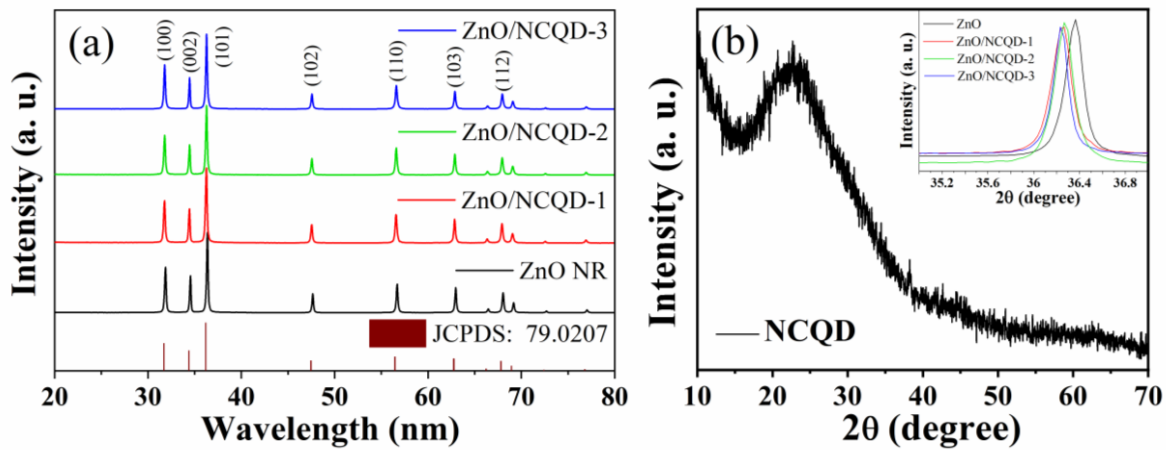

**Figure 1:** (a) XRD image of ZnO NR and ZnO/NCQD heterostructures, (b) XRD image of NCQD, inset shows the close looking image of the (101) peak of ZnO NR with the composites.

### 4.2 TEM analysis

To get a better picture of the formation of nanocomposites, the size and morphology of the as-synthesized samples have been carried out by transmission electron microscopy. Figure 2(a) shows a large area TEM image of as-synthesized pure ZnO with a rod-like morphology. The length of the ZnO nanorods is ~ (850 ± 50) nm and the width is ~ (120 ± 30) nm with monodispersive in nature. The closer view of two ZnO nanorods is depicted in Figure 2(b). High-resolution TEM (HRTEM) image of a single nanorod is presented in

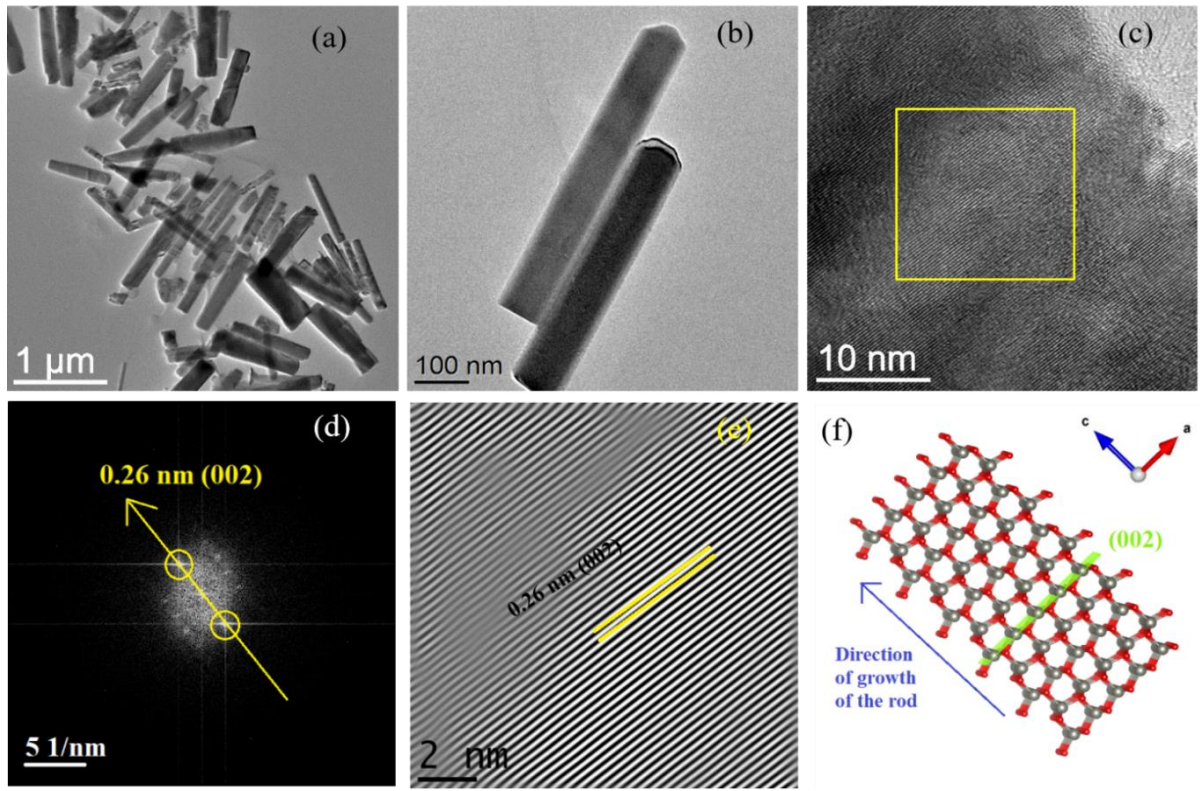

**Figure 2:** (a) Large area TEM image of ZnO NRs. (b) Closer view of two ZnO NRs. (c) HRTEM image obtained for a single nanorod. (d) FFT pattern obtained from the yellow square region of ZnO nanorod in (c) showing (002) plane of ZnO. (e) reconstructed HRTEM establishing the presence of (002) plane. (f) the atomic arrangement of ZnO showing (002) plane.

Figure 2(c). The first Fourier transform (FFT) pattern, obtained from the yellow-coloured box in Figure 2(c) represents (002) plane of ZnO with the corresponding *d* spacing of 0.26 nm (Figure 2(d)). The reconstructed HRTEM image in Figure 2(e) showcases the (002) plane of pure ZnO. Figure 2(f) reveals the atomic arrangement of the ZnO and the (002) plane has been displayed by green color. The characteristics of the wurtzite ZnO crystal structure is that, it consists of alternating planes composed of fourfold tetrahedrally coordinated $Zn^{2+}$ and $O^{2-}$ ions, stacked alternately along the c-axis (001). The {001} facets of ZnO are highly polar with high surface energy, which is the main factor for the preferential growth along the (001) direction. The capping agent PVP plays a major role in the formation of the rod-like structure [45]. In an aqueous solution, zinc cation ($Zn^{2+}$) from zinc nitrate, and, hydroxyl ion ($OH^-$) from HMT favorably react to form insoluble zinc hydroxide ($Zn(OH)_2$) quasi-precursor. This $Zn(OH)_2$ turns the solution cloudy milky in colour. Now the carbonyl functional groups (-C=O) of PVP coordinate with the $Zn^{2+}$ ions of ZnO crystal surfaces to form PVP– $Zn(OH)_2$ complex which controls the size and shape of the ZnO nanorods. The PVP– $Zn(OH)_2$ complex decomposes at 90 °C to form uniform ZnO nanorods.

The HRTEM image of as-synthesized NCQD is shown in Figure 3(a). The average size of NCQD lies between 2-3 nm. The large area TEM image of the as-synthesized NCQD decorated ZnO nanorods (ZnO/NCQD-2) are shown in Figure 3(b). A single NCQD decorated ZnO nanorod is unveiled in Figure 3(c) which depicts a clear boundary between the crystalline ZnO nanorods and amorphous NCQD. The amorphous nature of the NCQDs is also confirmed from the XRD analysis. An electrostatic attraction between the ZnO surface and NCQDs may play a crucial role in the formation of NCQD decorated ZnO NRs [52]. Further, HRTEM image of a nanorod is manifested in Figure 3(d) and the presence of NCQDs are indicated by the dotted boundary. The FFT pattern obtained from the yellow-colored box in Figure 3(d) is shown in Figure 3(e). The FFT pattern clearly reveals (002) plane of ZnO as obtained in the case of pure ZnO. Also, in this case, the calculated $d$ spacing corresponding the (002) plane is calculated as 0.26 nm, which is shown in the reconstructed HRTEM image (Figure 3(f)). We have also performed the EDAX and elemental mapping analysis of ZnO NR, ZnO/NCQD-1, ZnO/NCQD-2 and ZnO/NCQD-3 composite (see supporting file, Figure S1 to S4). The calculated atomic percentages are represented in Table S1 (supporting file). The occurrence of slightly higher concentration of O w.r.t Zn in EDAX analysis of bare ZnO NR is due to the presence of surface adsorbed oxygen on ZnO surface. Gradual increase of N doping is confirmed from ZnO/NCQD-1 to ZnO/NCQD-3 composites. In case of all composites, NCQD is present in an ample amount (see Table S1). This shielding of ZnO surface actually helps to reduce the photocorrosion effect of ZnO NR. Moreover, EDAX analysis shows that around 0.4 - 4.3 atomic % of N is contained within the composite materials. Elemental mapping analysis confirms the small but effective and uniform doping of nitrogen.

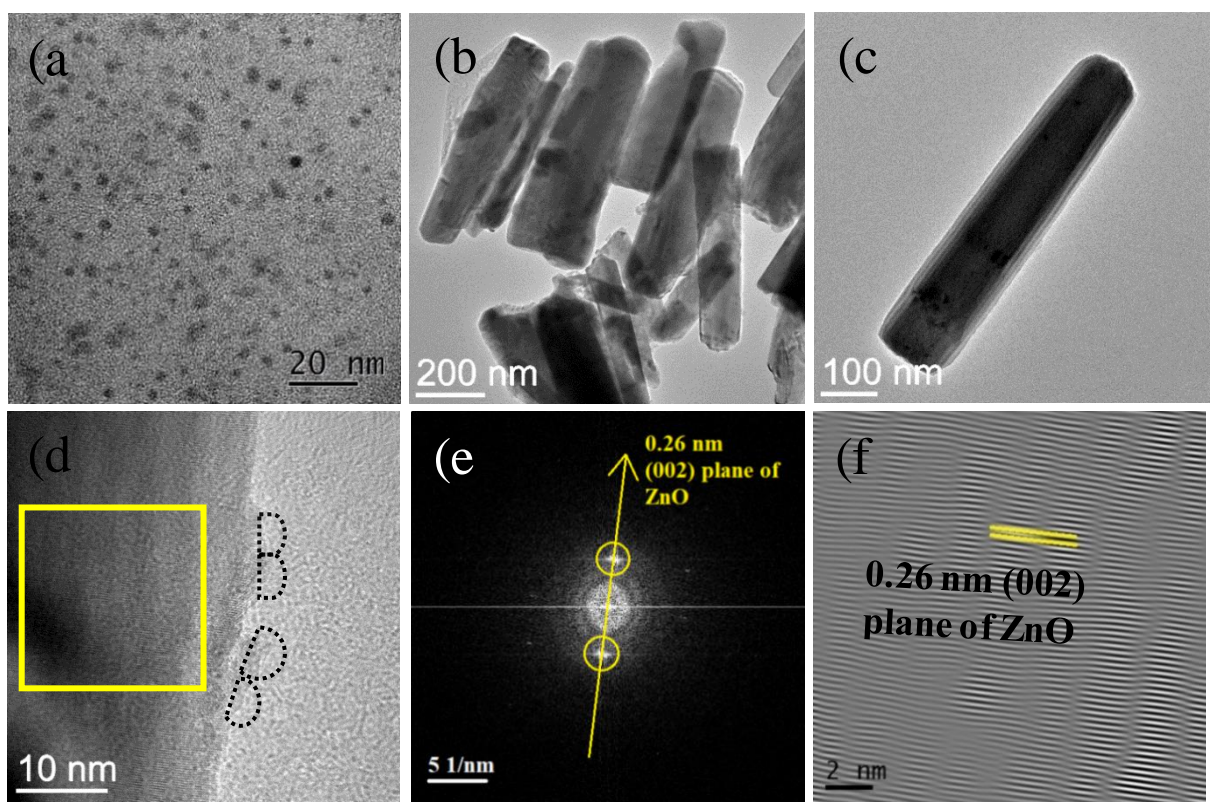

**Figure 3:** (a) HRTEM image of NCQD. (b) Large area TEM image of ZnO/NCQD-2 composite. (c) Closer view of a single NCQD decorated ZnO NR. (d) HRTEM image of NCQD decorated ZnO NR; the presence of NCQD is represented by the black dotted boundaries. (e) FFT pattern obtained from the yellow square region of ZnO nanorod in (d) showing (002) plane of ZnO. (f) reconstructed HRTEM confirming the presence of (002) plane.

### 4.3 XPS analysis

The XPS measurement has been used to study the elemental and surface properties of the ZnO/NCQD-2 composite. The full survey spectrum indicates the presence of zinc (Zn 2p), carbon (C 1s), and oxygen (O 1s) in the ZnO/NCQD composite as presented in Figure 4(a). From Figure 4(b), it can be observed that the binding energies of the Zn 2p components are slightly different. This is due to various surface morphologies of ZnO nanostructures [53]. The peaks at 1045.30 eV and 1022.20 eV are ascribed to binding energies of Zn $2p_{1/2}$ and Zn $2p_{3/2}$ respectively, which depicts a characteristic spin-orbit splitting of 23.1 eV. In Figure 4(c), the high-resolution peak of the C 1s spectrum at 285.40 eV is attributed to the C–C bond between $sp^2$ orbitals. In addition, the spectrum of O 1s (Figure 4d) in ZnO/NCQD-2 composite is divided into three peaks, the peak located at 530.72 eV is assigned to the core level of the crystal lattice $O^{2-}$ anions, while the broad peak at 531.86 eV is accredited to the O–H species or surface adsorbed water [54]. The inset of Figure 4(a) shows the high-resolution spectrum of N-related peak and the peak maxima around 400.30 eV confirms the presence of N in the ZnO/NCQD composite [55]. However, the XPS analysis shows that only a small amount of N is present in the ZnO/NCQD composite. This is not unexpected, because, during the formation of one-dimensional heterostructure only a small amount of NCQD have attached to the ZnO surface.

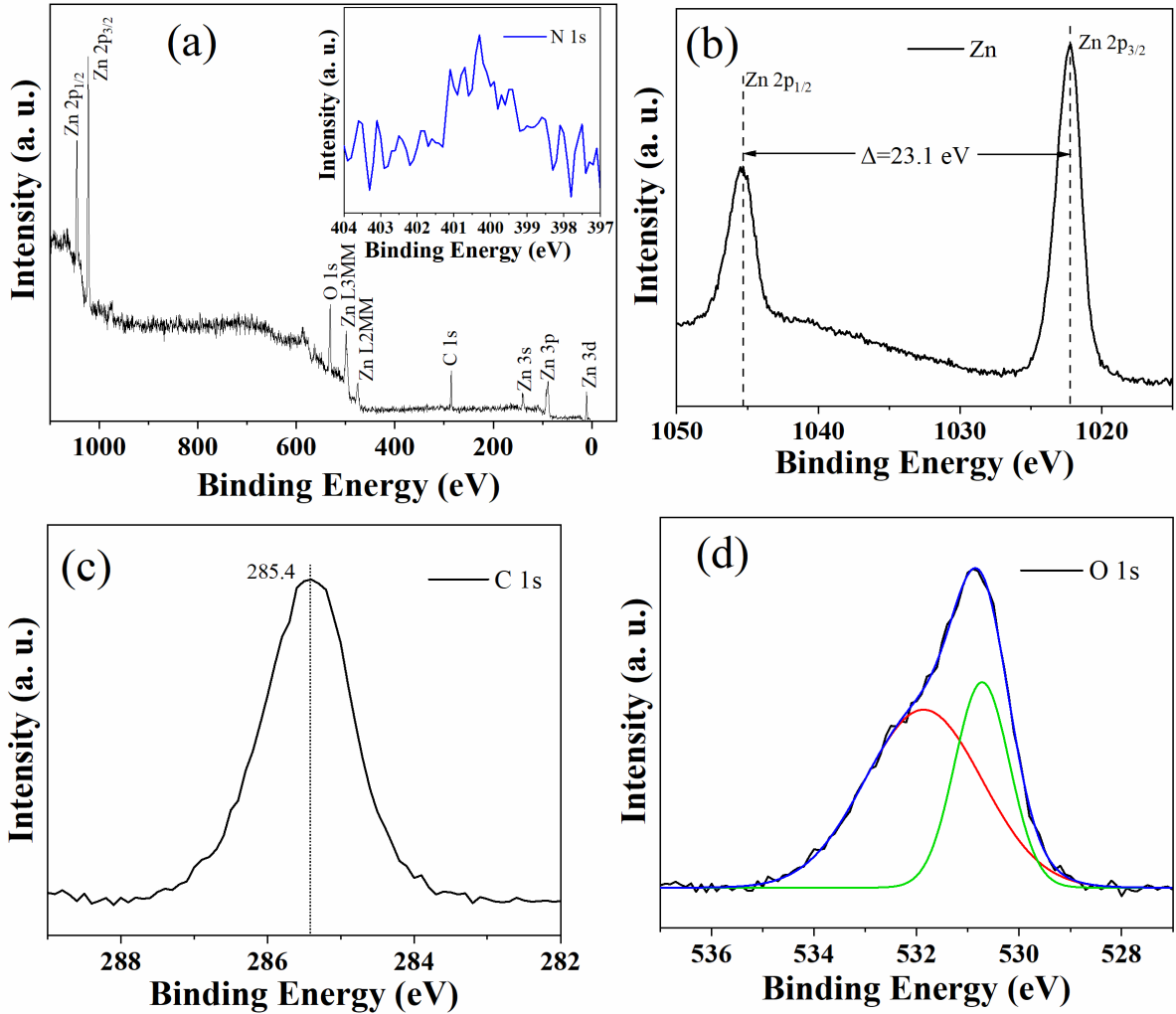

**Figure 4:** (a) XPS survey spectrum of ZnO/NCQD-2 composite, inset shows the spectrum of N 1s. (b), (c) and (d) show the XPS spectrum of Zn 2p, C 1s and O 1s respectively.

### 4.4 FTIR analysis

Figure 5 shows the FTIR spectra of NCQD, ZnO NR and ZnO/NCQD-2 composite. The characteristic peaks observed at 896 cm$^{-1}$ for ZnO NR, as well as, for the composite samples are due to the bending vibrating modes of Zn–O [56]. In the case of NCQD, the peak at 1048 cm$^{-1}$ can be ascribed to the C-O-C and C-O bonds, whereas, the peak at 1121 cm$^{-1}$ possibly occurs due to the C–O, C–N, and C–S bonds [57, 58]. The absorption band at 1225 cm$^{-1}$ in the case of ZnO/NCQD-2 originates from the stretching vibration of C-N which is absent in the case of pure ZnO [59]. This clearly indicates the formation of composite with NCQD. The peak at 1384 cm$^{-1}$ can be associated with the deformation modes of C-H, O-H and C–N groups [60]. The strong peak at 1657 cm$^{-1}$ in NCQD is due to the C=O vibration in CONH. Also, the identical C=O vibration in CONH has emerged in the case of ZnO and ZnO/NCQD composite at around 1645 cm$^{-1}$ [61]. The presence of this vibration mode in bare ZnO NR is because of the use of hexamethylenetriamine during synthesis. The observed minor peaks at 2852, 2924 and 2958 cm$^{-1}$ are due to the C-H bond stretching [58-62], while, the broad absorption band extending from 3000 cm$^{-1}$ to 3670 cm$^{-1}$ corresponds to O-H and N-H bonds [57, 59]. For further clarification, the FTIR spectra of these samples are displayed in the range 2500-4000 cm$^{-1}$ and deconvoluted them into three peaks and the deconvoluted spectra are shown in Figure S5(Supplementary file, Figure S5(a)-(c)). The deconvoluted image depicts that NCQD and ZnO/NCQD-2 samples contain a vibration mode at 3330 cm$^{-1}$ ascribed to the N-H bond, which is not prominent in pure ZnO NR. This clearly confirms the successful attachment of NCQD on the ZnO surface.

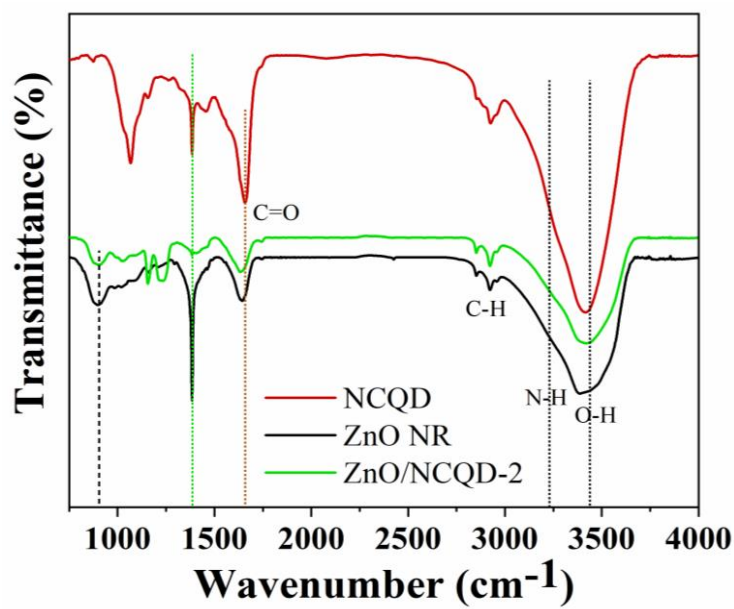

**Figure 5:** FTIR spectra of NCQD, ZnO NR and ZnO/NCQD-2 composite.

### 4.5 Optical studies

The optical absorption property of nanomaterials is an important tool to know about the bandgap. Besides, it can indicate the change of bandgap due to heterostructure formation, as well as predicts the photoactivation energy of the materials. The knowledge of these factors effectively helps prior to the study of photocatalytic activity.

The inset of Figure 6(a) represents the UV-vis absorption spectrum of NCQD. The spectrum of bare NCQD shows a shoulder at 272 nm due to the π−π* transition of the aromatic carbon (sp$^2$) [63], and the broad peak at around 322 nm is due to the n−π* transition of C=C, C=O, and C-N [64]. The optical properties of as-prepared ZnO NRs and ZnO/NCQD composites have been measured using UV-vis absorbance spectroscopy and depicted in Figure 6(a). Pure ZnO NRs shows a sharp absorption steep at ∼380 nm. At the optical band-gap absorbance edge, a slight redshift is observed for the composite samples compared to pure ZnO NR. This is due to the chemical interaction of NCQD with ZnO NRs. Moreover, ZnO/NCQD composites display a significant increase in absorption in the visible region. This can mainly be ascribed to the emergence of electronic interaction between the NCQD and ZnO [65, 66]. In addition, this absorption is more prominent in the case of ZnO/NCQD-2 composite. Hence, the result reveals the significant influence of carbon on the optical properties towards the improvement of visible-light absorption.

Also, from the absorption spectra, the band gap has been estimated by applying the Kubelka−Munk function as represented in Figure 6(b). The Kubelka−Munk function [F(R$_\infty$)] and photon energy (hν) can be estimated using equations 1-3 [67-69].

$$F(R_\infty) = (1 - R_\infty)^2/2 R_\infty \ldots\ldots\ldots (1)$$
$$R_\infty = 10^{-A} \ldots\ldots\ldots\ldots\ldots\ldots\ldots (2)$$
$$h\nu = 1240/\nu \ldots\ldots\ldots\ldots\ldots (3)$$

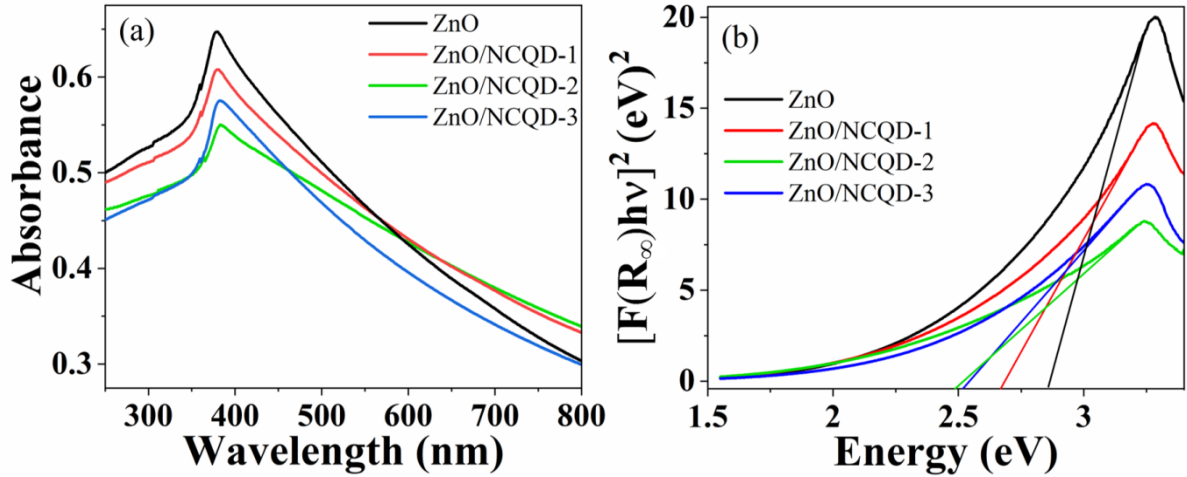

**Figure 6:** (a) UV–vis absorption spectra of ZnO NR and ZnO/NCQD composites, inset shows the UV-vis absorption spectra of NCQD. (b) the plot of the transformed Kubelka-Munk function as a function of the energy of light.

Here, $R_\infty$ is the reflection coefficient of the sample, A is the absorbance intensity, and $\lambda$ is the wavelength of the absorbed light. Though Kubelka-Munk function is originally meant for the reflectance spectra, it can be applied for absorption spectra through the Beer-Lambert Law as represented by Eq. (2) [68, 69]. Figure 6(b) shows the $[F(R_\infty)h\nu]^2$ versus $h\nu$ plot and the estimated bandgaps from the intercepts obtained from the straight-line fitting by the least square method are (2.77 ± 0.02), (2.69 ± 0.01), (2.44 ± 0.02) and (2.50 ± 0.01) eV for ZnO NR, ZnO/NCQD-1, ZnO/NCQD-2 and ZnO/NCQD-3 respectively. This is attributed to the sensitization of NCQD on the ZnO surface which effectively decreases the bandgap, more precisely, this is due to the formation of the type-II heterostructure. The presence of a surface defects and some functional groups generally introduces intraband gap states which reflects in broad absorption spectrum towards visible region, both in the case of bare ZnO NR and ZnO/NCQD composites. Because of this fact, the direct application of the Kubelka−Munk function plot underestimates the bandgap of the materials [68]. Also, we have calculated the bandgap values from Tauc plot [$(\alpha h\nu)^2$ vs $h\nu$], and, the obtained values are (2.54 ± 0.02), (2.37 ± 0.02), (2.15 ± 0.01) and (2.21 ± 0.01) eV for ZnO NR, ZnO/NCQD-1, ZnO/NCQD-2 and ZnO/NCQD-3, respectively. The $[(\alpha h\nu)^2$ vs $h\nu]$ plot is shown in the supporting file (Figure S6). In spite of having different methodological background, bandgaps calculated by the Kubelka-Munk function and Tauc plots are comparable, and, the issue of underestimation persists in both. Additionally, the effect of heterostructure formation on bandgaps can be understood from the discussion in the theoretical section.

Regarding photocatalysis, the photoluminescence (PL) study is an integral and essential tool to know about the charge dynamics like electron-hole pair recombination and the transfer of charges. The PL properties of the ZnO NR, NCQD and ZnO/NCQD composites have been investigated in detail. Figure 7(a) represents the PL spectra of ZnO NR, ZnO/NCQD-1, ZnO/NCQD-2 and ZnO/NCQD-3 with excitation at 300 nm. The near band edge emission (NBE) maximum of pure ZnO is at 381 nm, whereas, in the case of ZnO/NCQD-1, ZnO/NCQD-2 and ZnO/NCQD-3 composites it is at 382 nm, 384 nm and 388 nm respectively. Hence, slight red-shift has occurred due to the formation of the composites. Interestingly, due to the heterostructure formation of NCQD with ZnO NR, emission intensity has decreased significantly, indicating the presence of effective electronic interaction between ZnO and NCQD. The diminishing of PL intensity suggests a reduced recombination rate of photogenerated electron-hole pairs in the heterostructures. Formation of heterostructure with NCQD helps in efficient charge transportation in between ZnO and NCQD through the interface region. Interestingly, the lowest PL intensity is observed in case of ZnO/NCQD-2 composite. However, further increase in NCQD concentration increases the PL intensity, because excess NCQDs on ZnO act as recombination centres for the charge carriers [53]. This type of PL quenching suggests that ZnO/NCQD-2 composite ensures the most efficient charge transfer beneficial for efficient photocatalytic behaviour. Moreover, a slight change of PL emission profile of the composites than bare ZnO NR also confirms the presence of NCQD in the heterostructure. This is due to the synergic effect of emissions coming from ZnO NR and NCQD. Figure 7(b) shows the excitation-dependent PL spectra of NCQD. A red-shift in the PL peaks along with a change in intensity with increasing excitation wavelength is observed. This is consistent with previous literature reports and can be attributed to quantum size effect of NCQD [63, 70].

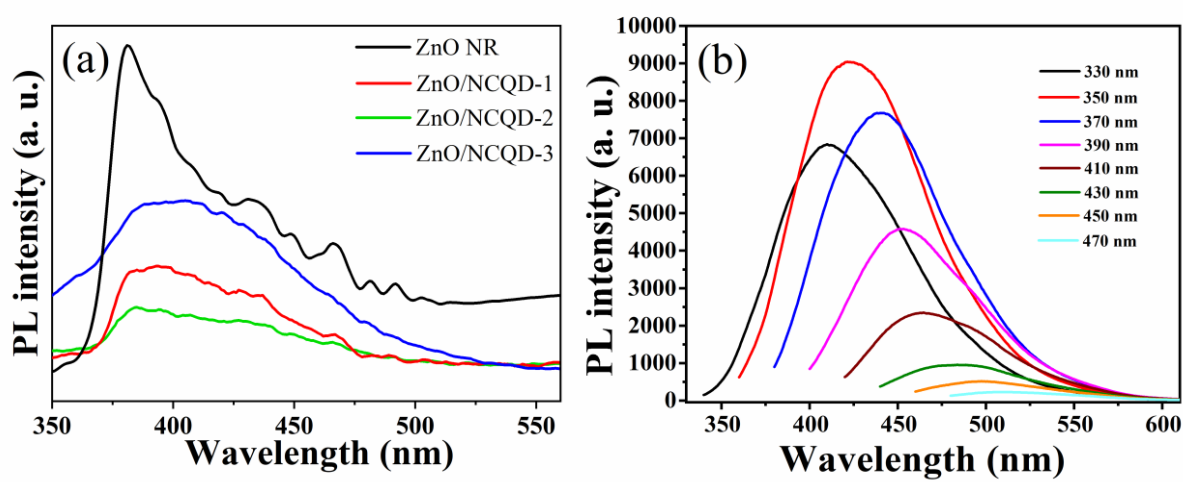

**Figure 7:** (a) The PL spectra of ZnO NR and NCQD decorated ZnO NRs, (b) PL spectra of NCQDs with different excitation wavelengths.

### 4.6 Photocatalytic study

We have taken RhB as a model dye to evaluate the photocatalytic efficiency of as-synthesized ZnO NR and their heterostructures with NCQD under UV radiation. The decrease in the concentration of the RhB dye with irradiation time by using different photocatalysts are shown in Figure S7(a)-S7(d) (Supplementary file). Figure 8(a) shows the $C/C_0$ versus irradiation time (t) plot, where, $C_0$ is the initial concentration of the dye at adsorption-desorption equilibrium and C is the remaining concentration of dye after a certain time of irradiation. The degradation efficiency can be calculated using the formula [71] as:

$$\text{Degradation (\%)} = (1 - C/C_0) \times 100\% \quad \ldots\ldots\ldots\ldots\ldots\ldots\ldots\ldots\ldots \text{(4)}$$

To study the self-photodegradability of RhB under UV light, the degradation reaction has been carried out in the absence of photocatalyst, and, the achieved percentage degradation of RhB is only 15% at the end of 30 min of irradiation as shown in Figure 8(a). We have investigated the photocatalytic activity of the synthesized samples up to 30 min of irradiation. It can be noted that bare ZnO NR shows good degradation efficiency and 69% dye degradation has achieved within 9 min of irradiation. However, in the case of ZnO/NCQD-2 almost 90% degradation has been achieved within this time of irradiation. After that, a very less amount of degradation ability has been observed. So, we can conclude that the maximum dye has degraded within 9 min of irradiation. The photocatalytic reaction kinetics corresponds to the Langmuir-Hinshlwood kinetic model and follows a pseudo first-order reaction process, which is expressed by $\ln(C/C_0) = k_{app}t$, where $k_{app}$ is the apparent reaction rate constant, C is the concentration of dye after irradiation time t and $C_0$ is the initial dye concentration [31]. The rate constants are obtained from the slope of the linear fitting of the logarithmic plot (Figure 8(b)). The calculated $k$ values for ZnO NR, ZnO/NCQD-1, ZnO/NCQD-2 and ZnO/NCQD-3 are $(12.5 \pm 0.2) \times 10^{-2}$, $(14.3 \pm 0.5) \times 10^{-2}$, $(20.8 \pm 0.7) \times 10^{-2}$ and $(12.9 \pm 0.5) \times 10^{-2}$ min$^{-1}$ respectively. So, ZnO/NCQD-2 turns out to be the photocatalyst with best degradation efficiency among all the composites. Further increase in NCQD amount however reduces the adsorption of dye in ZnO surface resulting decrease in efficiency. In fact, excess amount of NCQD shades the open surface of ZnO. As a result, interaction of dye with ZnO decreases. Moreover, the photocatalytic activity of the NCQD is weaker than that of ZnO. Thus, excess amount of NCQD decoration on the ZnO NR creates a competition between the light absorption capacities of ZnO NR with NCQD, which leads to a decrease in photocatalytic efficiency [52].

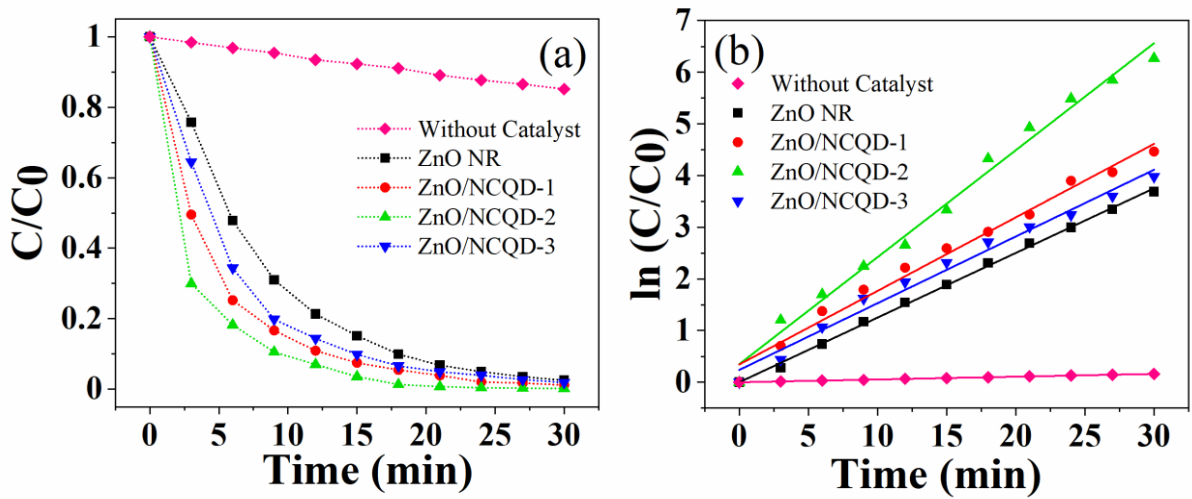

**Figure 8:** (a) Photocatalytic degradation curve ($C/C_0$ vs. time) and (b) $\ln(C_0/C)$ versus time plot for RhB dye degradation using ZnO NR and ZnO/CQD composites.

**Mechanism of photocatalytic activity:**

Generally, the effectiveness of a photocatalytic process depends on three major facts- (i) generation of electron-hole pairs (ii) separation of electron-hole pairs i.e., reduction of electron-hole recombination and (iii) photo-oxidation and photo-reduction reactions on the active surfaces of the catalyst [72]. Here, NCQD is decorated on ZnO surface in a small amount. The major electron-hole pair generation comes from the ZnO surface. So, we can conclude that generation of electron-hole pair in case of composites is similar to bare ZnO NR. However, due to the presence of some C=O, C=N groups on the NCQD surface, the composite can attract the cationic dye RhB on the catalyst facilitating photo-oxidation and photo-reduction reactions. But this can only be done to a small extent. Hence the major reason for enhanced photocatalytic activity comes from the separation of electron-hole pairs. The composite of ZnO with NCQD forms a type-II heterostructure, which is later explained in the theoretical section. Under irradiation electrons and holes are created in the conduction band (CB) and valence band (VB) of both ZnO and NCQD respectively. Photoexcited electrons can travel from NCQD CB to ZnO CB due to the more negative CB position of NCQD and holes can travel in the opposite direction from the more positive VB of ZnO to NQCD (See Figure 13 for detail). This leads to a complete electron-hole separation, thus, enhancing the photocatalytic behavior which is evident from the significant decrease in PL intensity. Photogenerated electrons can reduce dissolved oxygen into superoxide radicals ($\dot{O}_2^-$). This highly oxidizing radical however can be transformed into hydrogen peroxide ($H_2O_2$), hydroperoxyl radical ($H\dot{O}_2$) and hydroxyl radical (OH·). On the other hand, holes ($h^+$) react with hydroxyl ions to form hydroxyl radicals. During photocatalytic progression, these highly reactive radicals and free electrons/holes react with absorbed RhB molecules and degrade this completely into mineral acids, $CO_2$, and $H_2O$. The overall photocatalytic reaction processes can be represented as follows:

$$ZnO/NCQD + h\nu \rightarrow ZnO\ (e^-) + NCQDs\ (h^+) \quad \ldots\ldots (5)$$
$$e^- + O_2 \rightarrow \dot{O}_2^- \quad \ldots\ldots\ldots (6)$$
$$2H^+ + \dot{O}_2^- + \dot{O}_2^- \rightarrow H_2O_2 + O_2 \quad \ldots\ldots (7)$$
$$\dot{O}_2^- + H_2O \rightarrow H\dot{O}_2 + OH^- \quad \ldots\ldots (8)$$
$$H\dot{O}_2 + H\dot{O}_2 \rightarrow OH\cdot + O_2 \quad \ldots\ldots (9)$$
$$H\dot{O}_2 + H_2O_2 \rightarrow H_2O + O_2 + OH\cdot \quad \ldots\ldots (10)$$
$$h^+ + H_2O/OH^- \rightarrow OH\cdot \quad \ldots\ldots (11)$$
$$OH\cdot/\dot{O}_2/h^+ + RhB\ dye \rightarrow CO_2 + H_2O \quad \ldots\ldots (12)$$

The formation of active species like OH· radicals during the photocatalytic process can be detected by using the fluorescence technique, where terephthalic acid (TA) has been used as a probe molecule. This is one of the effective approaches to compare the photocatalytic activity of different catalysts worldwide [71]. Here, we have tested the photocatalytic efficiency of ZnO NR and ZnO/NCQD composites. TA reacts with hydroxyl (OH·) radical to generate 2-hydroxyterephthalic acid (TAOH) which has a fluorescence peak at about 425 nm on excitation at 315 nm. Fluorescence intensity of TAOH is directly proportional to the amount of OH· radical produced on the surface of the photocatalysts during the UV light illumination. The fluorescence intensity ratio ($I_t/I_0$) of TAOH generation with respect to irradiation time, monitored at 425 nm is represented in Figure 9. Where, $I_0$ is the intensity of TAOH without irradiation, and, $I_t$ is the intensity of TAOH after certain time of irradiation. The intensity ratio shows that ample amount of OH· radicals are generated in case of ZnO/NCQD-2 composite compared to the other materials. This experiment reveals that ZnO/NCQD-2 composite generates sufficient amount of OH· radicals and shows effective photocatalytic activity.

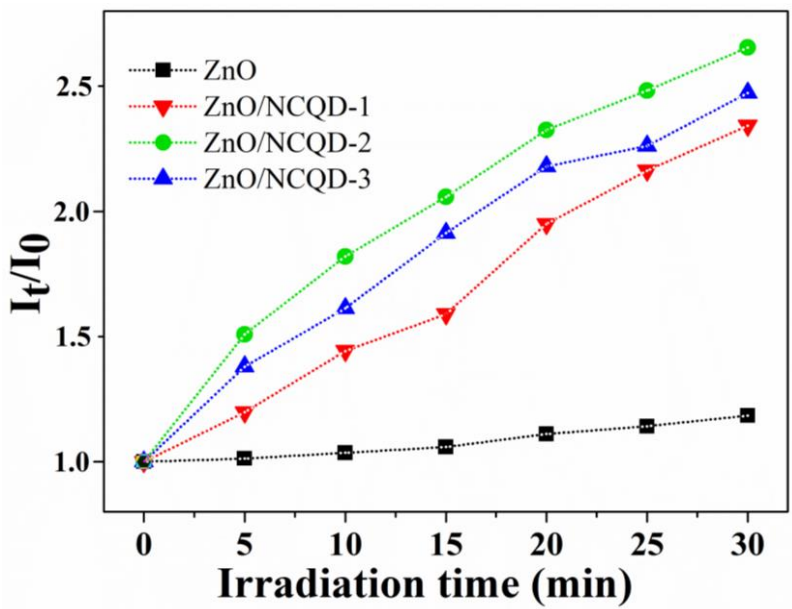

**Figure 9:** The photoluminescence intensity ratio ($I_t/I_0$) of 2-hydroxyterephthalic acid with irradiation time for ZnO NR and ZnO/NCQD composites.

The reusability and stability of the photocatalysts are of utter importance in view of feasible photocatalytic applications. The photocatalytic repeatability of the best composite (ZnO/NCQD-2) along with bare ZnO NR have been investigated via a cycling test up to 21 min of irradiation. For each recycling test, the catalysts are collected, washed with a water-ethanol mixture and finally dried at 70 °C. Figure 10(a) shows the removable concentration of dye after each cycle run. In the case of bare ZnO NR, the degradation efficiency of RhB decreases from 90% to 72% after five consecutive cycling runs, whereas, a lesser decrease in degradation efficiency have been observed in the case of ZnO/NCQD-2 (99% to 95%). The reduction in degradation efficiency of ZnO/NCQD composites is probably due to the mass loss percent of photocatalyst during each cycle. The result indicates that after the formation of heterostructure with NCQD, the catalyst becomes more stable against photocorrosion. This is due to the shielding effect of NCQD which prevents ZnO against photocorrosion [73]. We have also recorded the XRD spectrum of ZnO/NCQD-2 composite after five successive reuse cycles and represented in Figure 10(b). The result reveals that XRD spectrum has not changed significantly after recycling experiments. This observation further confirms the high photostability and anti-photocorrosion nature of the ZnO/NCQD-2 composite photocatalyst.

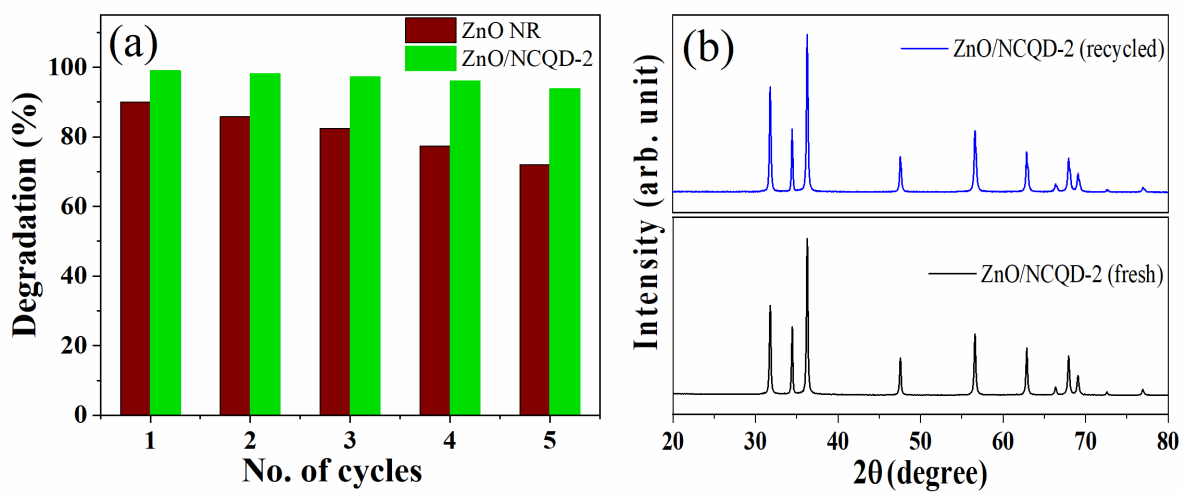

**Figure 10:** (a) Recycle test of ZnO NR and ZnO/NCQD-2 composite, (b) XRD spectra of ZnO/NCQD-2 synthesized (colour Black) and after 5$^{th}$ recycling test (colour blue).

### 4.7 Photocurrent study

To further confirm the electron-hole charge transfer process under photoexcitation, the photocurrent measurement of pure ZnO NRs and ZnO/NCQD-2 composite have been performed. Figure 11(a) and (b) respectively depict the UV light illuminated ($\lambda$= 360 nm) I-V characteristic curves of ZnO NR and its optimized composite ZnO/ NCQD-2, along with the dark currents. The dark current in ZnO/NCQD-2 is significantly higher than bare ZnO NR. This indicates the improvement of the conductivity of ZnO NR by NCQD. Photocurrent in ZnO/NCQD-2 is almost 10 times higher than bare ZnO NR. Thus, in heterostructure, the NCQD effectively helps in reducing the electron-hole pair recombination. As a result, photogenerated electrons are accumulated in the CB of ZnO and take part in electrical conduction. This increased conductivity of the composite increases the photocurrent densities. The effective charge transfer across the heterojunction interface of ZnO/NCQD-2 increases the photocurrent densities which is also evident from the improved photocatalytic activity.

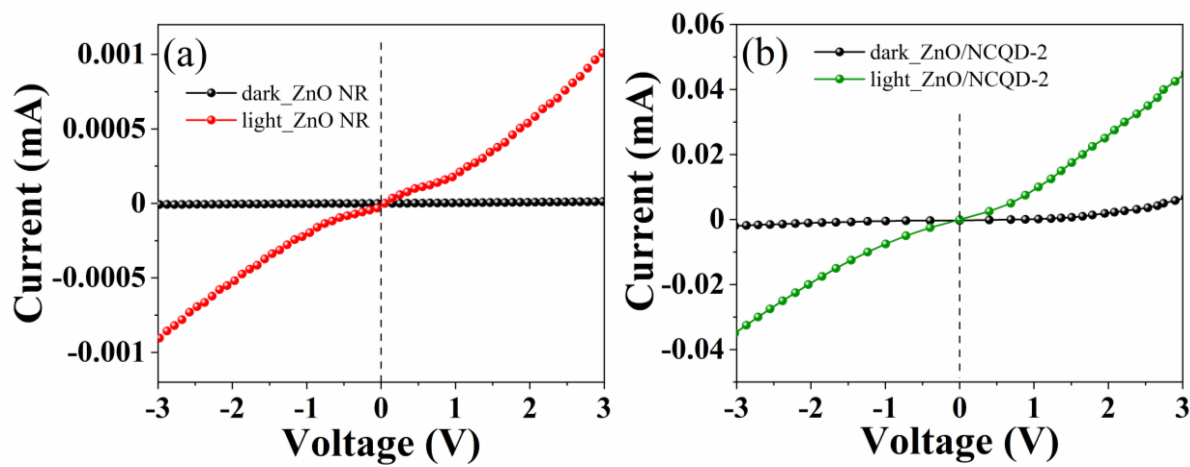

**Figure 11:** Current-voltage (I–V) characteristics curve of (a) ZnO NR and (b) ZnO/NCQD-2 in dark along with the presence of UV irradiation respectively.

### 4.8 Cyclic Voltammetry study

To investigate the influence of the loading of NCQDs upon ZnO nanorods, a cyclic voltammetry study has been performed. The CV study gives us an effective method to know further about the charge transfer mechanism [74]. The cyclic voltammograms of the synthesized samples have been displayed in Figure 12. The inset of Figure 12 shows the enlarged image of cyclic voltammograms of positive potential region. The reduction peak currents (cathodic peak current) of ZnO NR, ZnO/NCQD-2 and ZnO/NCQD-3 are 0.26, 0.49 and 0.39 mA respectively. The increased amount of current compared to pure ZnO NRs in case of ZnO/NCQD-2 suggests a higher number of electrons have been transformed to ZnO from NCQD [75]. Therefore, the differences in the flow of current are related to the different electron transfer at the interface of the electrodes which reflects in the difference in charge densities of different sets of the composites. This implies a greater electron–hole separation in the ZnO/NCQD-2 composite [76]. Due to this increased electron-hole separation, ZnO/NCQD-2 gives the best photocatalytic activity as well as better photoconductivity. However, excess loading of NCQDs in the ZnO/NCQD-3 composite reduces the reduction peak current as compared to ZnO/NCQD-2. This indicates that the charge transfer is decelerated in this case resulting in the reduction of the photocatalytic activity. Thus, the CV study gives us an overall idea about the charge-transfer process in the synthesized composites.

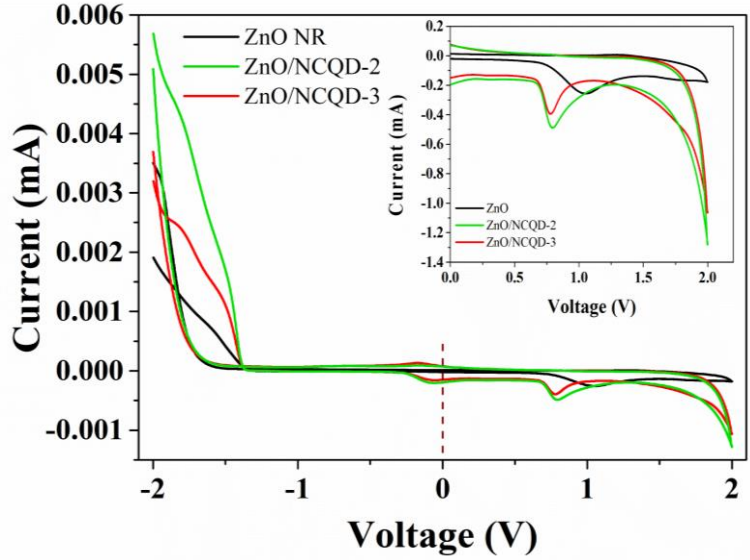

**Figure 12:** Cyclic voltammograms of ZnO NR, ZnO/NCQD-2 and ZnO/NCQD-3 respectively.

### 4.9 Theoretical investigations

To complement our experimental observation, we move further towards a theoretical understanding of the key factors for this artificial photocatalysis. There are two different parameters governing the photocatalytic performance of semi-conductors. The first one is its signature property, the bandgap. The thermodynamic voltage for water splitting at normal temperature and pressure (NTP) is 1.23 V [77], however, due to energy loss, a photo-electrochemical cell with an illuminated electrode of bandgaps greater than 1.60 eV is required [78].

But the bandgap unilaterally cannot determine the photocatalytic profile of any semiconductor. The band alignment plays a pivotal role as well. In normal hydrogen electrode (NHE) scale, the valence band maxima (VBM) should be more positive than the water oxidation level ($E_{H_2O/O_2}$=1.23, 0.81 V for pH = 0, 7), whereas the conduction band minima (CBM) should be more negative than the hydrogen production potential ($E_{H^+/H_2}$= 0, -0.41 V for pH = 0, 7) [79]:

$$H_2O + 2h^+ \rightarrow 2H^+ + \frac{1}{2}O_2$$
$$2H^+ + 2e^- \rightarrow H_2 \ldots\ldots\ldots\ldots (13)$$

The theoretical methodology to determine the VBM-CBM is not straightforward but a little tricky one. It is essential to calculate the work function (WF) to proceed further. The work function is the energy required to liberate an electron from the surface of any piece of material which equals to the difference between its Fermi energy ($E_F$) and the vacuum potential energy ($V_{vacuum}$).

WF = $V_{vacuum}$ - $E_F$ ……………. (14)

The knowledge of bandgap and WF is enough to locate VBM and CBM energies. For WF calculation, to mimic the semi-infinite system with one open surface, the slab is created with three layers of ZnO along (002) direction and a 12 Å vacuum is created to make the system aperiodic along the direction. As the WF is the difference of vacuum energy for slab and the Fermi energy of the bulk ($WF = V_{vacuum} - E_F^{Bulk}$), hence, two different calculations, one on bulk ZnO and another on the slab have to be carried out [80]. As the electrostatic potential within the interstitial region of the material, irrespective of slab or bulk, should be the same, matching these two values would standardize the energy levels [81]. The difference of the macroscopic averages for bulk and slab ($\Delta_V = V_{bulk} - V_{slub}$) works as a correction to $E_F^{Bulk}$ to bring equivalence of energy levels of these two separate self-consistent calculations [82].

$$E_{F,corrected}^{Bulk} = E_F^{Bulk} - \Delta_V \ldots\ldots\ldots\ldots\ldots (15)$$
$$WF = V_{vacuum} - E_{F,corrected}^{Bulk} \ldots\ldots\ldots\ldots\ldots (16)$$

The lattice constants of the relaxed structure of ZnO using PBE exchange-correlation is found as 3.2689 and 5.2779 Å following the hexagonal Wurtzite P6$_3$mc space group, in which Zn and O occupy (0.333333, 0.666667, 0.000000) and (0.333333, 0.666667, 0.378948) sites, respectively.

But, to appropriately locate the band edges or to find the bandgap of semiconductors using the DFT method, we have to overcome another hurdle. The local density approximation and most of the generalized gradient approximations underestimate the bandgap [83]. For oxides, the strong correlation between electrons makes the situation more challenging and that is why it is still tricky to manage ZnO theoretically. We have used PBE0 hybrid-functional method to estimate the band gap and band alignments [84]. The calculated bandgap of ZnO is 3.15 eV, which is comparable to the experimental observation. The band structure and the density of states of ZnO is represented in Figure S8 in the supplementary file.

Now, following the method for calculating the work-function described above, we locate the Fermi energy for ZnO (002) surface at -5.95 eV with respect to the vacuum energy. Hence, the VBM is at -7.52 eV, which matches well with the experimentally observed ionization potential (IP) [85, 86]. Similarly, the CBM -4.37 eV is in agreement with the experimental electron affinity (EA) (-4.20 to -4.52 eV) [87, 88] for (002) surface.

The bandgaps of CQDs are size and shape-dependent [89, 90]. Here to understand the mechanism we have taken four different CQDs of different shapes as shown in Figure S9 (Supplementary file). The rectangular CQD (28 atoms) has the smallest bandgap 1.13 eV, followed by the triangular (22 atoms) having 2:05 eV and hexagonal (24) CQD (24 atoms) 4.23 eV. All the structures have hydrogen passivation at the edges. The structural relaxation using PBE functional has revealed that the hexagonal structure is most stable. So, we focus only on the hexagonal structure, and, for other CQD structures, refer to Figure S9. As the experiment is on NCQD decorated ZnO system, therefore, in the next step, we replace a carbon atom with hydrogenated nitrogen (N-H) as shown in Figure 13(a) as NCQD. We have provided the atomic coordination of CQD and NCQD in the supplementary information for better understanding of the structures. The bandgap of the relaxed hexagonal NCQD reduces to 2.94 eV.

Now, as we see in Figure 13(a), the CBM of the ZnO surface does not qualify the $E_{H^+/H_2}$ = 0, -0.41 V vs. NHE condition for the complete range of pH = 0 to 7. This is why a formation of heterostructure with other elements becomes necessary, and, when NCQD/ZnO heterostructure is formed, the necessary condition is indeed satisfied. Through this finding, justification of the performance enrichment of ZnO/ NCQD heterostructure finds firm footing. There can be surfaces of different shapes and sizes but the general physics behind that is simple which we unfold here through this ab-initio calculation.

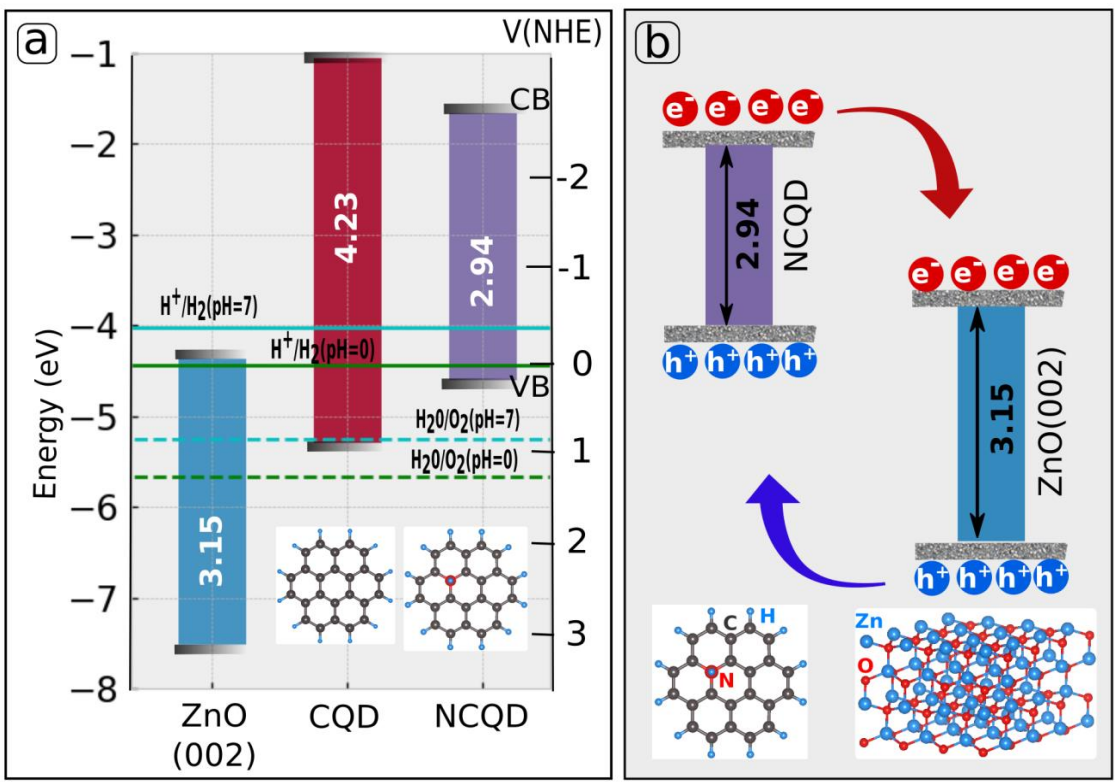

**Figure 13:** (a) The band alignment of ZnO (002) surface compared with hexagonal different carbon quantum dots (CQD) and nitrogenated CQD (NCQD). (b) Schematic diagram of Type-II heterostructure formation between ZnO and NCQD.

Now, is there any reason behind our claim on better performance of ZnO/NCQD composite over the previously reported ZnO/CQD ones? Figure 13(a) is self-explanatory on this issue. The hexagonal CQD has higher bandgap (4.23 eV) than the hexagonal NCQD (2.94 eV). On the other hand, relative to the vacuum energy, the CBM of ZnO-002, CQD and NCQD are placed at -4.37 eV, -1.06 eV and -1.67 eV, respectively. While producing heterostructure, the CBM barriers, which is the energy difference of the CBM levels of components forming the heterostructure, are calculated as 2.70 eV for ZnO/NCQD and 3.31 eV for ZnO/CQD. As the CBM barrier of ZnO/NCQD is less than ZnO/CQD, hence, less energy is needed for the reaction at CBM. So, we see that the CBM position is the reason behind the improvement of the efficiency of ZnO/NCQD over ZnO/CQD. The schematic formation of the ZnO/NCQD heterostructure is depicted in Figure 13(b). The diagram indicates a Type-II heterostructure formation. Electrons moving from NCQD CBM to ZnO CBM initiate hydrogen evolution reaction (HER), and, the transfer of holes from ZnO-VBM to NCQD-VBM can release the Oxygen. In addition, previous reports endorse that N-doped graphene systems show better conductivity than pristine ones [91, 92]. Efficient charge transfer is observed in our experiments as well. Summing up all these, we can readily understand the reason for the efficiency enrichment of ZnO/NCQD over ZnO/CQD or bare ZnO for RhB degradation.

## 5. Conclusion

The goal of the human race should be to provide nature-friendly technological advancement in a cost-effective way. Here, one-dimensional ZnO NR, NCQD and their composites have been synthesized via a cost-effective soft chemical route to fulfill the requirement of the degradation of RhB dye, a significant water pollutant. The XRD analysis have revealed the intrinsic amorphous nature of the NCQDs, while, the HRTEM images confirm the nanorod formation of ZnO along the (002) direction with a length of ~ (850 ± 50) nm and a width of ~ (120 ± 30) nm. Using EDX analysis, we have found that the NCQDs significantly covers the ZnO nanorods. Due to the formation of such ZnO/NCQD heterostructures, the bandgaps of the composites have effectively reduced. The photoluminescence quenching in composites indicates that the NCQDs have played a key role in the charge transfer process. Charge transfer mechanism has further been analyzed using cyclic voltammetry study, which suggests that the increased electron-hole separation in ZnO/NCQD-2 is the reason behind its better photocatalytic and photoconducting efficiency than others. Moreover, recycle test reveals that ZnO/NCQD composites are less vulnerable under photocorrosion, as, a part of the ZnO surface is covered by the NCQDs. However, we have also found that higher concentration of NCQDs can shield the ZnO nanorods, thus, reducing the efficiency. It indicates that an optimum choice of NCQD concentration is the enabling key factor for the ZnO/NCQD composites.

Interestingly, under UV irradiation, all these composites have shown excellent photocatalytic degradation capability of RhB compared to the bare ZnO NRs. Theoretical investigations have revealed that the formation of Type-II heterostructure between ZnO and NCQD have generated photoinduced electron transfer from the CBM of NCQD to CBM of ZnO NR. And, a simultaneous photoinduced hole transfer has occurred from the VBM of ZnO nanorod to the VBM of NCQD, resulting in complete charge separation. This effectively suppresses the electron-hole recombination, which reflects in efficacious photocatalytic activity. Finally, we can conclude that the combination of salient features, like the higher absorption of UV light, better charge transfer, less photocorrosion, and, an effective type II heterostructure formation contribute to achieve efficacy in the degradation of RhB (~90% within nine minutes) using the sustainable and cost-effective ZnO/NCQD composites. This work provides useful information on the fabrication of NCQD decorated catalysts, which may pave a way for futuristic catalyst designing towards new energy sources and pollutant eradication.

**Acknowledgements**
The author S. K. Mandal sincerely acknowledge CSIR (NET Fellowship), India and S. Paul acknowledge SERB (National Post Doctoral Fellowship), India for providing the fellowship during the tenure of the work.

**Credit Author Statement**
S. K. Mandal has synthesized the materials and has carried out the photocatalytic study through XRD, FTIR, UV-Vis spectroscopy, PL spectroscopy, Photocurrent and CV measurements. S. Paul have analyzed the TEM and CV portion of this manuscript. S. K. Mandal, S. Paul and D. Jana have analyzed the experimental part of this manuscript. S. Datta have done the theoretical calculations and analysis. S. Mandal and S. Datta have primarily prepared the experimental and theoretical parts of the manuscript, respectively. All have discussed and further edited the whole manuscript. As per our knowledge, there is no conflict of interest.

**Supplementary Information**
Figure S1, S2, S3 and S4 represents the EDS data of ZnO NR, ZnO/NCQD-1, ZnO/NCQD-2 and ZnO/NCQD-3 composite respectively. Figure S5(a), (b) and (c) represent the deconvoluted FTIR image of NCQD, ZnO NR and ZnO/NCQD-2 composite. Figure S6

represents the Tauc plot of ZnO NR and ZnO/NCQD composites. Figure S7: (a), (b), (c) and (d) represents the UV–Vis absorption spectra of rhodamine B dye using 5 mg of bare ZnO NR, ZnO/NCQD-1, ZnO/NCQD-2 and ZnO/NCQD-3 respectively as a catalyst with UV light irradiation up to 30 min. Figure S8 represents (a) Energy band diagram and (b) Density of States (DOS) of ZnO calculated using PBE0 hybrid functional. The calculated band-gap is 3.15 eV at Γ. Figure S9 represents the VBM and CBM alignment of CQD of different shapes and ZnO (002) layer. The bandgaps calculated using PBE0 hybrid functional are depicted on the bar diagrams.